# Source of Energetic Protons in the 2014 September 1 Sustained Gamma-ray Emission Event

N. GOPALSWAMY*

*NASA Goddard Space Flight Center, Greenbelt, MD*

P. MÄKELÄ, S. YASHIRO, S. AKIYAMA, H. XIE, N. THAKUR

*The Catholic University of America, Washington, DC*

*Corresponding author. E-mail: nat.gopalswamy@nasa.gov

Abstract

We report on the source of >300 MeV protons during the SOL2014-09-01 sustained gamma-ray emission (SGRE) event based on multi-wavelength data from a wide array of space- and ground-based instruments. Based on the eruption geometry we provide concrete explanation for the spatially and temporally extended γ-ray emission from the eruption. We show that the associated flux rope is of low inclination (roughly oriented in the east-west direction), which enables the associated shock to extend to the frontside. We compare the centroid of the SGRE source with the location of the flux rope's leg to infer that the high-energy protons must be precipitating between the flux rope leg and the shock front.  The durations of the SOL2014-09-01 SGRE event and the type II radio burst agree with the linear relationship between these parameters obtained for other SGRE events with duration ≥3 hrs. The fluence spectrum of the SEP event is very hard, indicating the presence of high-energy (GeV) particles in this event. This is further confirmed by the presence of an energetic coronal mass ejection (CME) with a speed >2000 km/s, similar to those in ground level enhancement (GLE) events. The type II radio burst had emission components from metric to kilometric wavelengths as in events associated with GLE events. All these factors indicate that the high-energy particles from the shock were in sufficient numbers needed for the production of γ-rays via neutral pion decay.

## 1. Introduction

Gamma-ray events temporally extended beyond the impulsive phase of solar flares were first reported by Forrest et al. (1985) using the Solar Maximum Mission's Gamma Ray Spectrometer (SMM/GRS) data. During the 1982 June 3 event, the γ-ray emission extended beyond the flare impulsive phase by ~20 min and was recognized as the neutral-pion decay continuum based on different spectra during the impulsive and late phases. The Gamma-ray burst experiment (PHEBUS) on board the GRANAT mission also observed extended-phase emission at energies >10 MeV (Talon et al. 1993; Vilmer et al. 2003). Using data from Gamma-1 telescope, Akimov et al. (1991) reported on the 1991 June 15 event with the extended phase γ-ray emission lasting for more than 2 hr. Based on data from the Energetic Gamma Ray Experiment Telescope (EGRET) on board the Compton Gamma Ray Observatory (CGRO), Kanbach et al. (1993) reported on another event with a duration exceeding 8 hr. After the advent of the Fermi Large Area Telescope (Fermi/LAT, Atwood et al. 2009) it has become clear that such extended-duration γ-ray events are rather common (Ackermann et al. 2014; Share et al. 2018; Winter et al. 2018). To underline the fact that γ-ray photons are emitted long after the associated flares, the emission is now referred to as sustained γ-ray emission (SGRE) (see e.g., Plotnikov et al., 2017; Klein et al. 2018; Gopalswamy et al. 2018a; Kahler et al. 2018). As pointed out by Ryan (2000), the definition of the long-duration γ-ray events has been imprecise: the term "long-duration gamma-ray flare (LDGRF)" refers to the γ-rays as flare, although the flare is gone long before the end of the γ-ray events.

During SGRE events, the photon spectrum extends to energies >1 GeV and has a peak around 70 MeV, characteristic of γ-rays from neutral pion decay. It was already recognized that the extended γ-ray emission is related to the process accelerating SEPs that is distinct from the impulsive phase acceleration (Forrest et al. 1985). The definite evidence that large SEP events are produced by shocks driven by coronal mass ejections (CMEs) was first reported by Kahler et al. (1978), soon after the discovery of CMEs in white light (Tousey 1973) (see Reames, 1999 for a review). Therefore, the idea of shocks supplying the necessary >300 MeV protons has also been proposed soon after the discovery of the long duration γ-ray flares (Murphy et al. 1987; Ramaty et al. 1987). Nevertheless, the idea of impulsive-phase particles trapped in long loops and precipitating slowly to the photosphere continues to be pursued (e.g., Hudson 2018; Grechnev et al. 2018; de Nolfo et al. 2019a,b).

Citing the presence of metric type II burst, Akimov et al. (1991) suggested that the protons responsible for the γ-ray emission should have been shock-accelerated. The 2.223 MeV γ-ray line (GRL) emission observed by SMM/GRS from a backside eruption on 1989 March 29 located ~10º behind the limb (Vestrand and Forrest 1993; Cliver et al. 1993) was shown to be consistent with a shock source for the energetic protons. The 2.223 MeV line is produced deep in the chromosphere, so it would not reach the observer on the Sun-Earth line from behind the limb. However, particles accelerated at the front of the associated shock can readily supply particles precipitating on the frontside of the Sun to produce the line emission (Cliver et al. 1993; Vestrand and Forrest, 1993). Cliver et al. (1993) provided details of the CME that was driving the shock and the spatial extent of the CME/shock was consistent with the spatially extended GRL emission. The shock acceleration was further corroborated by the presence of a metric type II radio burst, which is indicative of the shock near the Sun. Cliver et al. (1993) were cautious to state that it was an open question whether shock mechanism applicable to GRL emission would also apply to SGREs. The observation of SGRE from three backside events by Fermi/LAT provided definite evidence that the time-extended emission is also spatially extended (Pesce-Rollins et al. 2015a,b; Plotnikov et al. 2017; Ackerman et al. 2017; Jin et al. 2018). In particular, the SOL2014-09-01 (Sep14, for short) event was extraordinary in that the eruption was ~ 40º behind the limb, yet SGRE was observed by Fermi/LAT suggesting that the SGRE emission is truly spatially extended and must be due to a CME-driven shock.

Share et al. (2018) noted that most of the SGRE events are associated with a fast CME (>800 km/s) and a type II burst at decameter-hectometric (DH) wavelengths. One of the most recent revelations is that the duration of SGRE events is linearly related to that of DH type II bursts and inversely related to the ending frequency of the type II bursts (Gopalswamy et al. 2018a). These observations provide the physical basis for the time-extended nature of SGRE events: life time of such SGRE events is determined by the duration over which the underlying CME-driven shock efficiently accelerates >300 MeV protons that propagate to the solar surface and produce γ-ray emission via the decay of neutral pions. Since there is a one-to-one correspondence between large SEP events and DH type II bursts extending from metric to kilometric wavelengths (Gopalswamy et al. 2008), it is highly likely that the same shock is responsible for >300 MeV protons (for SGRE) and ~10 keV electrons (for DH type II burst). It must be noted that the shock continues to accelerate particles to lower energies (e.g., energetic storm particle events are often observed after the end of SGRE) and weak/fragmented type II bursts are observed until the shock arrives at the observing spacecraft such as Wind and STEREO.

Several papers have already appeared dealing with various aspects of the Sep14 event, but here we focus on the source of >300 MeV protons that provides a concrete explanation for the spatially and temporally extended γ-ray emission from a solar eruption. We show that the SGRE duration and type II burst duration are consistent with the linear relationship found by Gopalswamy et al. (2018a). Furthermore, we exploit the two-view observations Solar Terrestrial Relations Observatory (STEREO, Kaiser et al. 2008) and the Solar and Heliospheric Observatory (SOHO, Domingo et al. 1995) to derive the flare, flux-rope, and shock structures in the eruption that are consistent with the idea of shock particles precipitating on the frontside of the Sun to produce the observed SGRE event.

## 2. Data Description

Sep14 is the most intense and longest-lasting among the three behind-the-limb SGRE events observed by Fermi/LAT (Pesce-Rollins et al. 2015a,b; Share et al. 2018; Ackermann et al. 2017; Plotnikov et al. 2017; Jin et al. 2018; Hudson 2018; Grechnev et al. 2018). Plotnikov et al. (2017), Ackermann et al. (2017), and Share et al. (2018) described the full time evolution of the SGRE flux. Share et al. (2018) showed that the SGRE event was distinct from the impulsive phase emission. Details of the eruption region and the impulsive-phase radio and X-ray emissions have been already described in these papers. Here we describe additional data that help fully understand the eruptive event and the SEP event that led to the SGRE. Ground-based radio instruments observed radio bursts of various types: type II, type III, and type IV continuum (Carley et al. 2017). The radio bursts also continued into the interplanetary (IP) medium as observed by the Radio and Plasma Wave Experiment (WAVES, Bougeret et al. 1995) on board the Wind and STEREO (SWAVES, Bougeret et al. 2008) spacecraft. We make use of the fact that the type II radio burst was observed by SWAVES without obstruction, so we obtain the full evolution of the burst in the radio dynamic spectrum. The early phase of the shock has already been described in previous papers (Plotnikov et al. 2017; Jin et al. 2018; Grechnev et al. 2018). Here we provide information on the shock and CME as observed by the Large Angle and Spectrometric Coronagraph (LASCO, Brueckner et al. 1995) on board SOHO and the Sun Earth Connection Coronal and Heliospheric Investigation (SECCHI, Howard et al. 2008) on board STEREO. More details on the DH type II burst can be found in https://cdaw.gsfc.nasa.gov/CME_list/radio/waves_type2.html. CME properties measured in the sky plane are available online in the SOHO/LASCO CME catalog (https://cdaw.gsfc.nasa.gov, Yashiro et al. 2004; Gopalswamy et al. 2009a). To get the three-dimensional speeds, we fit a flux rope to the CME observed in coronagraph and EUV images using the graduated cylindrical shell (GCS) model (Thernisien 2011) and a spheroid to the shock ahead of the CME (Olmedo et al. 2013; Hess & Zhang 2014; Mäkelä et al. 2015; Xie et al. 2017; Gopalswamy et al. 2018b). Presence of >300 MeV protons is a critical requirement for the SGRE events. We use GOES >100 MeV proton data available from NOAA's Space Weather Prediction Center (SWPC) as a proxy to the >300 MeV protons. For particles arriving at STEREO spacecraft, we use data from the High Energy Telescope (HET) and Low Energy Telescope (LET) of the In Situ Measurements of Particles and CME Transients (IMPACT, von Rosenvinge et al. 2008). The highest-energy channel of the STEREO particle detectors is ~100 MeV. In order to get >100 MeV integral fluxes similar to GOES, we extrapolate the proton spectrum to ~1 GeV and assume that there is little contribution from particles at energies >1 GeV. We also obtained the fluence spectrum of the event from STEREO and GOES data.

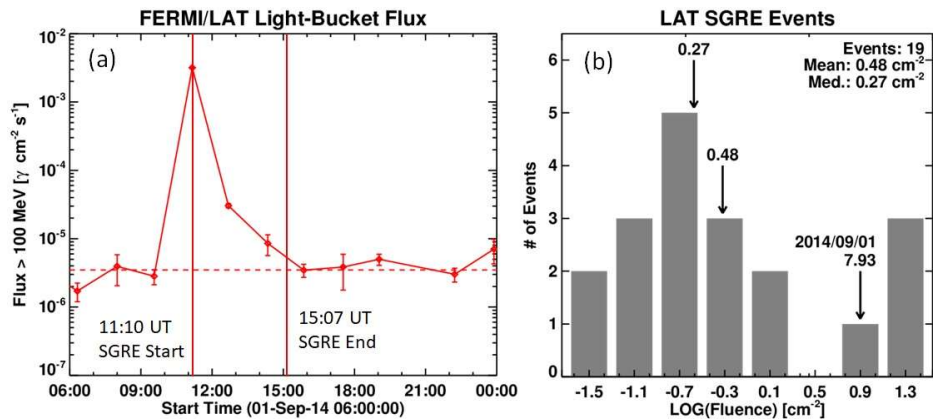

Figure 1. Time profile of the Sep14 SGRE event (a) and its fluence compared with the >3hr SGRE events observed by Fermi/LAT (b). The duration over which the fluence was computed is shown in (a). The horizontal dashed line in (a) represents the celestial and solar quiescent backgrounds within a 10-degree circle around the Sun. The Sep14 event is the fourth largest among the 19 SGRE events that have durations exceeding 3 hr.

## 3. Analysis and Results

We define the SGRE duration as the interval from the end of the impulsive phase as marked by the GOES soft X-ray peak and the midpoint between the last signal data point above the background level and the one after that (Gopalswamy et al. 2018a). Note that Fermi/LAT observes the Sun only intermittently due to the continuously changing pointing direction of the instrument. The SGRE peak occurred after the impulsive phase of the flare as revealed by the hard X-ray emission observed by the High Energy Neutron Detector (HEND) of the Gamma-ray Spectrometer onboard the Mars Odyssey mission (see Grechnev et al. 2018 for details). Therefore, we estimate the duration of the event from the SGRE peak at 11:10 UT to 15:07 UT as 3.92±0.76 hr. Figure 1 shows the SGRE time profile and the fluence compared with those of other SGRE events of duration >3 hr. The Sep14 event had the fourth largest fluence (7.94 cm$^{-2}$). Some of the γ-ray photons would not have reached Fermi/LAT, so the true fluence of the Sep14 event might have been higher. The three events with higher fluence are: 2012 March 7 (24.1 cm$^{-2}$), 2014 February 25 (14.9 cm$^{-2}$), and 2017 September 10 (13.9 cm$^{-2}$). The fluence values reported here are slightly different from those in Winter et al. (2018) because their start times are different from ours. For example, in the Sep14 event, we use the peak time of the soft X-ray emission (11:10 UT) as the starting time, whereas Winter et al. (2018) use 11:02 UT, which is in the impulsive phase.

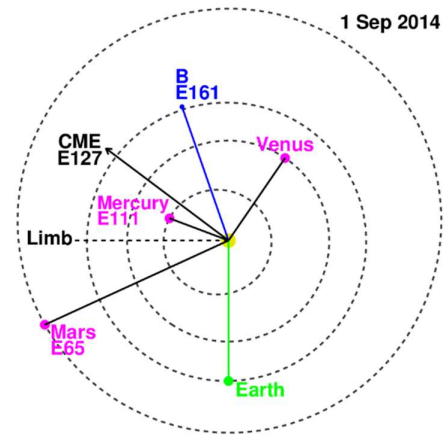

Figure 2. Locations of STEREO B (B), Mercury, and Mars with respect to Earth at the time of the Sep14 event. The CME direction (E127) is shown. The CME was front-sided to STEREO-B, Mercury MESSENGER, and Mars Odyssey. The line above the east limb of the Sun from Earth view is marked. For STEREO-B, the eruption is a western event; for MESSENGER and Mars Odyssey it is an eastern event.

## 3.1 The solar source

The Sep14 SGRE event originated from an active region located ~37º behind the east limb (N14E127 – see Ackermann et al. 2017; Plotnikov et al. 2017; Jin et al. 2018; Grechnev et al. 2018). Fortunately, there were three space observatories in which the source was observed as a frontside event (see Fig. 2). STEREO Behind (STB) spacecraft was located at E161, so the solar source location is N14W34 in that view. STB's Extreme Ultra-violet Imager (EUVI) imaged the eruption region including the coronal dimming and post-eruption arcade (PEA), while COR1 and COR2 imaged the CME from the inner corona to the IP medium. STB was also well connected to the SEP event. The hard X-ray emission from the flare was observed by HEND onboard the Mars Odyssey mission (see Grechnev et al. 2018 for details). Mars Odyssey was located at a longitude of E62, so the eruption was a disk event (~E65) in the spacecraft's view. Mercury MESSENGER was at a longitude of E110, so the eruption was a disk-center (E17) event in its view. The Solar Assembly for X-rays (SAX), which is the Sun-pointed detector of the X-Ray Spectrometer (XRS) on board MESSENGER detected the soft X-ray flare (see Share et al. 2018 for details). The SGRE source location was estimated as N41E90 by Ackermann et al. (2017). However, the source location has been revised and the new location is closer to the latitude of the eruption site (M. Pesce-Rollins, private communication). In any case, the SGRE source longitude is W71 in STB view, which is about 37º west of the active region.

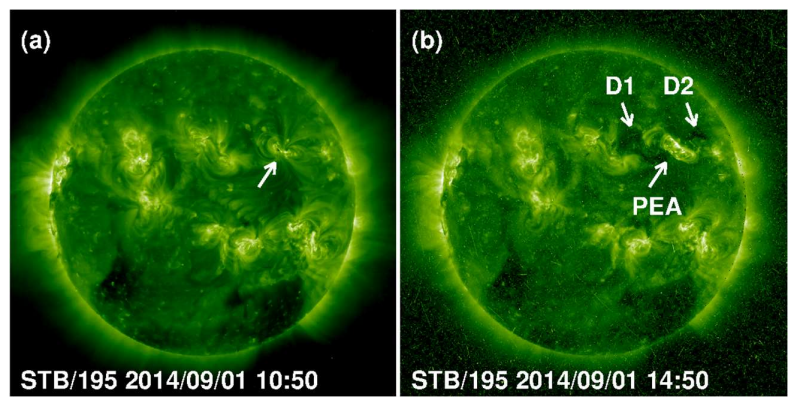

Figure 3. An overview of the Sep14 eruption as seen in STB/EUVI images taken just before the eruption (b) and in the post-eruption phase (b). In (a), the arrow points to the eruption region. In (b), the two dimming regions D1, D2 and the post-eruption arcade (PEA) are marked. The "snowstorm" in the image in (b) is due to energetic particles from the eruption hitting the STB detectors.

The STB/EUVI images in Fig. 3 show the region of interest before and after the eruption. The eruption resulted in a PEA oriented along a NE-SW line and a twin dimming (D1, D2). When the active region rotated onto the disk a few days later (September 5 onwards), it was found that the leading polarity was positive. The dimming regions D1 and D2 had negative and positive polarities, respectively. The inclination of the arcade was ~-34º, while the line joining the dimming regions was almost horizontal (-12º inclination).

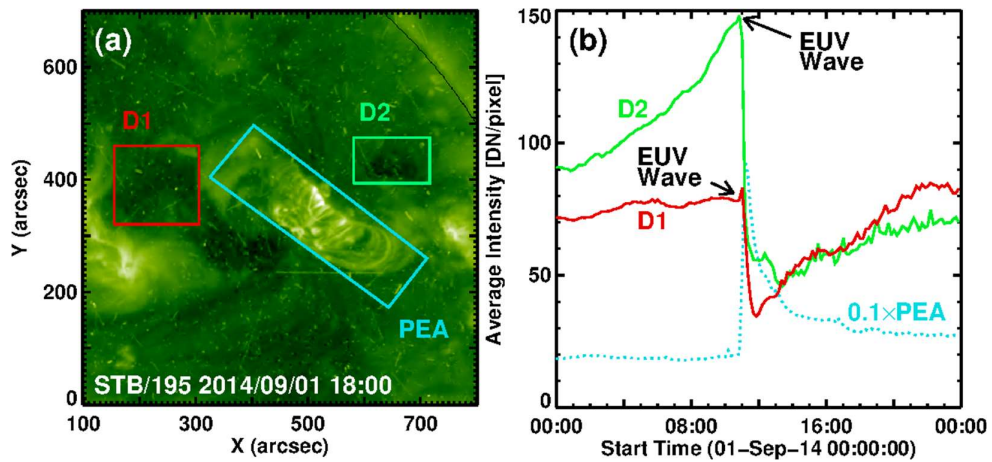

Figure 4. The core dimming regions D1, D2, and the PEA of the Sep14 eruption (a) and the time evolution of the intensities (b) in those regions. The small spike in D1 and D2 correspond to the arrival of the EUV wave above these regions. The intensity of the PEA (blue dotted line) is equivalent to the soft X-ray flare. There was additional dimming to the southwest of D1 caused by the removal of an overlying structure due to the eruption. This dimming did not recover as D1 and D2 did. The concerned active region rotated on to the disk as NOAA AR 12158. The region produced an eruption on 2014 September 10. The PEA at that time was more east-west than in the present event. The box D2 was fixed while the sun was rotating underneath. Some bright structures in the outer part of the active region rotated into the box, increasing the signal slowly (over ~12 hours), unrelated to the eruption.

Figure 4 shows the evolution of the EUV intensity in the dimming regions D1 and D2 and the PEA. The spikes in the dimming region plots correspond to the EUV wave dome passing these regions before the dimming starts. D1 and D2 are also referred to as core dimming and thought to be the locations where the CME flux rope footpoints are rooted (Webb et al. 2000; Dissauer et al. 2018; Gopalswamy et al. 2018b). The deepest dimming occurred slightly before 12 UT, but there was a small reversal in D2. The recoveries were similar in D1 and D2 for several hours. The EUV intensity of the PEA peaks around 11:10 UT, coincident with the soft X-ray peak observed by SAX/XRS onboard MESSENGER (Share et al. 2018). The equivalent soft X-ray size was estimated using the similarity (Nitta et al. 2013; Chertok et al. 2015) between EUV and X-ray time variations as X2.4 (Ackermann et al. 2017).

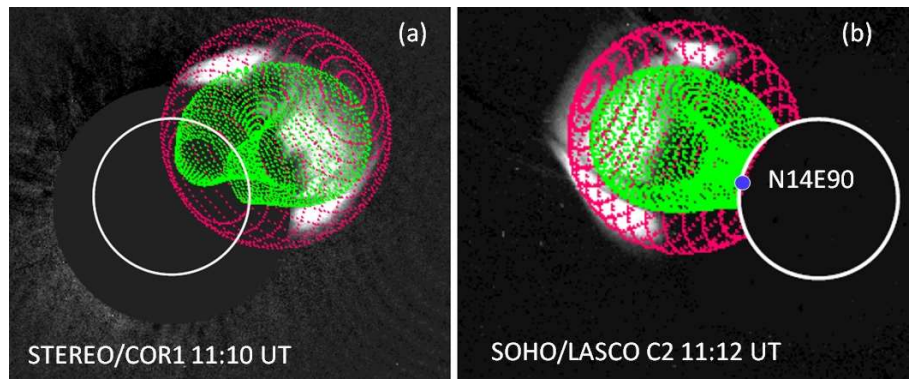

Figure 5. Flux rope (green) and shock (pink) obtained from the forward-modeling fit to the CME from STEREO and SOHO observations. The flux rope and shock are shown superposed on STB/COR1 image at 11:10 UT (a) and SOHO/LASCO/C2 image at 11:12 UT. The blue dot on the east limb is the SGRE centroid obtained from Fermi/LAT. The location corresponds to the outskirts of the western leg of the flux rope within the shock sheath as seen in STB view. The SGRE flux reached its peak value around the time of these images. The shock and flux rope leading edges are at 3.2 and 2.8 Rs, respectively. The GCS fit also gives the ratio of the flux rope radius to the heliocentric distance of the flux rope nose as 0.24.

## 3.2 The CME and shock

Mass motion from the eruption region was observed in SDO images starting at 10:58 UT as reported by Grechnev et al. (2018), with a leading-edge height at 1.35 Rs at 11:00:01 UT. The CME was also observed by GOES SXI at 11:01:15 UT with the leading edge at a height of 1.7 Rs (Carley et al. 2017). Both observations were in sky-plane projection. In order to get the three-dimensional speed of the CME and shock, we used STEREO and SOHO observations to fit a flux rope to the CME and a spheroid to the leading shock dome using the GCS model for the flux rope (Thernisien 2011) and the spheroidal model for the shock (Olmedo et al., 2013). The flux rope-shock model has been applied successfully to track the leading edges of CME events (Hess and Zhang, 2014; Mäkelä et al., 2015; Xie et al., 2017).

Figure 5 shows the shock and flux rope fits to STEREO and SOHO coronagraph images at 11:10 UT and 11:12 UT, respectively. The direction of propagation of the flux rope nose is along N16E117, which corresponds to a slight westward deflection of the flux rope compared to the radial direction (N14E127). The deflection brings the nose closer to the limb (~27º behind the limb in Earth view). The tilt angle of the flux rope is -12º, which is consistent with the east-west placement of the core dimming regions D1 and D2 (see Fig. 4) and different from the tilt of the PEA (~ -34º). The flux rope's face-on and edge-on half widths are 38º and 18º, respectively. Recall that the angular distance from the source region to the west limb in STB view is ~37º. This means the western edge of the flux rope extends to the frontside of the Sun. The shock, which is more extended than the flux rope, clearly extends to the frontside and is consistent with the EUV wave crossing the STB west limb and propagating to the frontside (Plotnikov et al. 2017; Jin et al. 2018). In the case of high-inclination flux ropes, the east-west extent would be smaller and a source at 40º behind the limb may not produce a γ-ray event on the Earth-facing disk. The inclination difference was previously invoked by Gopalswamy et al. (2015a) to explain why the 2014 January 6 SEP event was a GLE event (low-inclination flux rope), while the 2012 May 27 SEP was not (high-inclination flux rope) even though the eruption longitudes were similar behind the west limb.

Figure 6 shows the height, speed, and acceleration of the shock and flux rope as a function of time within the STB/COR2 field of view (FOV). Both the shock and flux rope have average speeds exceeding 2000 km/s. The shock speed attains a peak value of ~2450 km/s at 11:25 UT and then decrease slowly due to the aerodynamic drag but remains above 2000 km/s within COR2 FOV. The speeds and initial accelerations are typical of CMEs producing GLEs in SEP events. For comparison, the CME speed is in between that of the two GLE events of solar cycle 24: 2012 May 17 (~2000 km/s; Gopalswamy et al. 2013a) and 2017 September 10 (~3400 km/s; Gopalswamy et al. 2018c). The initial acceleration (1.66 km s$^{-2}$) is similar to that of the 2012 May 17 event (1.77 km s$^{-2}$), but much smaller than that of the 2017 September 10 event (9.1 km s$^{-2}$). The CME and shock kinematics are thus consistent with an energetic eruption capable of accelerating particles to GeV energies.

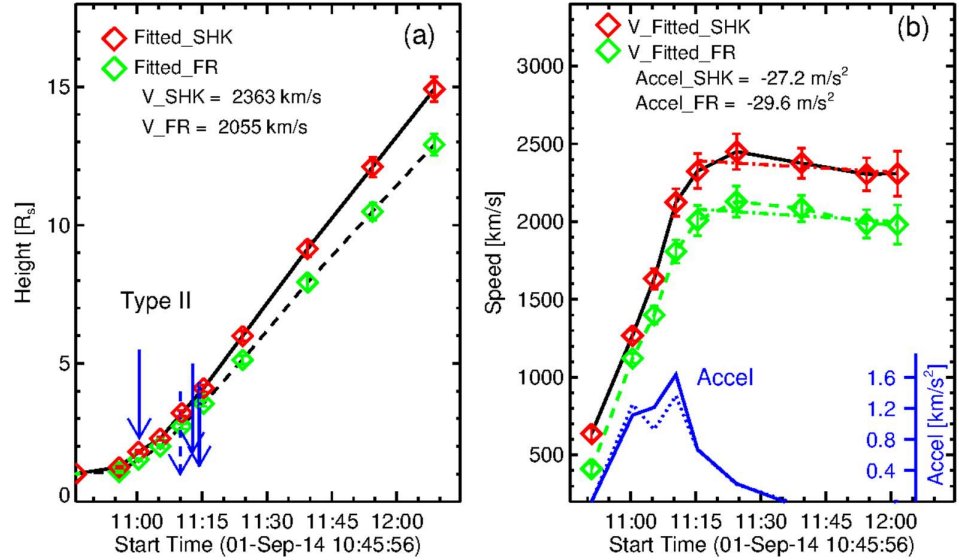

Figure 6. Kinematics of the Sep14 flux rope (green data points and dashed line) and shock (red data points and solid line) based on the GCS and spheroidal models. (a) height-time plot of the shock and flux rope. The onset times of metric (solid arrows) and DH (dashed arrow) type II bursts are marked. (b) The evolution of shock and CME speeds and the acceleration of shock (solid blue line) and flux rope (dashed blue line) derived from the height-time data points. A peak shock acceleration was 1.6 km s$^{-2}$ attained at 11:10 UT. The dot-dashed lines are best-fit lines to the decreasing speeds of the flux rope and shock.

### 3.3. Radio bursts

The Sep14 SGRE event was accompanied by complex radio emission from microwave to kilometer wavelengths indicating the presence of accelerated electrons. Details of the metric and microwave emission have been reported in Fig. 1 of Carley et al. (2017) and in Fig. 9 of Grechnev et al. (2018). Here we provide a comprehensive description including the IP components. The IP components are particularly important, because they indicate continued particle acceleration required for the SGRE event. Figure 7 shows a composite dynamic spectrum in the range 500 MHz to 10 kHz. The data above 100 MHz are from the Orfées and CALLISTO spectrographs. The Nancay Decametric Array (NDA) data were used in the range 10-88 MHz. At frequencies below 16 MHz, we have used the SWAVES data from STB. The type IV emission above ~200 MHz was interpreted as moving type IV burst by Carley et al. (2017) due to ~MeV electrons trapped in the CME flux rope structure. These frequencies are not occulted because the flux rope (and hence the moving type IV source) is already above the limb at the start of the emission (~11:01 UT) as evident from the height of the soft X-ray transient (~1.69 Rs at 11:01:15 UT). Intense metric radio emission starts at <180 MHz. The emission consists of a series of type III bursts, three episodes of type II bursts, and a flare continuum.

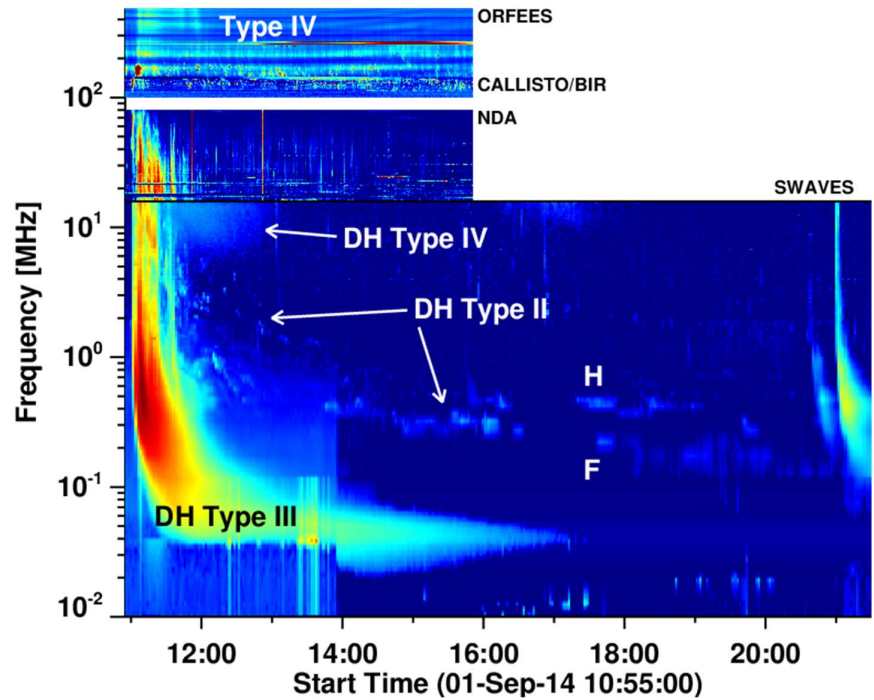

Figure 7. Radio emissions associated with the Sep14 SGRE event shown in the form of a composite dynamic spectrum. Observations at frequencies >100 MHz are from Orfées and CALLISTO instruments. Nancay Decametric Array (NDA) observes in the range 10-88 MHz at both polarizations, but here we have shown the total intensity. The sharp start of radio emission at ~180 MHz is most likely due to the occultation of plasma levels >90 MHz (the radio emission observed at the limb is typically at harmonic). Decameter-hectometric (DH) Type II, type III and type IV emissions are marked. The bright emission in the NDA band continues as the long duration type IV. The DH type II has fundamental-harmonic (F-H) structure.

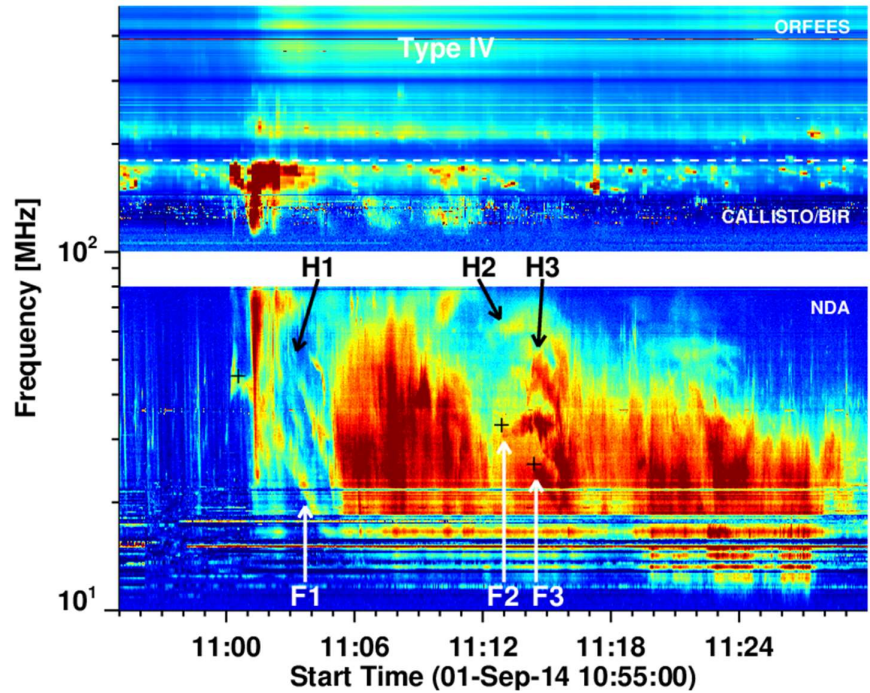

Figure 8. Composite dynamic spectrum showing nonthermal radio emission at frequencies >10 MHz. The plus symbols denote the starting time/frequency of the three type II episodes with the fundamental and harmonic components denoted by F1, F2, F3 and H1, H2, H3. The horizontal dashed line at 180 MHz denotes the high-frequency cutoff of the plasma emissions. The type IV burst at frequencies above 200 MHz is gyrosynchrotron emission from the CME flux rope.

Carley et al. (2017) fitted a gyrosynchrotron spectrum to the higher-frequency radio observations that peaked at ~1000 MHz and inferred a magnetic field strength of 4.4 G at a height of ~1.3 Rs. This value is consistent with the range of axial field strengths (10 to 100 mG at 10 Rs) of coronal flux ropes obtained by Gopalswamy et al. (2018d). A magnetic field of 4.4 G at 1.3 Rs corresponds to an axial field strength of ~75 mG at 10 Rs, assuming self-similar expansion of the flux rope. This is above the average of the distribution (52 mG) because the eruption is very energetic: higher total reconnected flux in the eruption region results in higher axial field strength and larger CME kinetic energy (Gopalswamy et al. 2018d). It is also well known that the axial field strength and CME speed are correlated (Gonzalez et al. 1998; Gopalswamy et al. 2015b).

In the DH domain, the eruption was accompanied by type III, type II, and type IV bursts. The DH type III bursts were a continuation of type III bursts in the NDA frequency range and lasted until ~11:25 UT. The DH type II burst crosses the upper edge of the dynamic spectrum (~16 MHz) around ~11:10 UT and continues as fragmented emission until ~20:00 UT with an ending frequency below 200 kHz. However, the burst has a clear break during 16:15 to 17:30 UT. Type III bursts typically mark electron acceleration in the impulsive phase of the associated flare and propagation along open field lines, while the type IV indicates electrons trapped in tall flare structures, i.e., PEAs (Gopalswamy 2011). Type IV bursts at frequencies below 10 MHz are very rare and are associated with CMEs with very high average speed (~1500 km/s). The low-frequency edge of the type IV burst descends to lower frequencies, reaching a minimum of ~7 MHz at 12 UT and then increases, reaching the upper edge of the SWAVES frequency range an hour later. The type II burst is at the descending edge of the type IV burst, which is typically the case when type II and type IV bursts occur together. While the type II and type III bursts are also observed by Wind/WAVES, the type IV burst was not because of its directivity (Gopalswamy et al. 2016a). It must be noted that the presence of type II emission from metric to kilometric domains is a characteristic of GLE events indicating the presence of ~GeV protons.

At lower frequencies (<100 MHz), the radio emission lasted for about 30 min and contains several components: type III, type II, and a flare continuum. The flare continuum is weaker than the type II and type III bursts. The flare continuum is thought to extend to lower frequencies as DH type IV continuum (Gopalswamy et al. 2016a). Three episodes (1, 2, 3) of type II bursts are observed, and whose fundamental (F1, F2, F3) and harmonic (H1, H2, H3) components are marked in Fig. 8. The starting times and frequencies of the fundamental components (F1, F2, F3) are shown in Table 1.

**Table 1.** Type II burst episodes from NDA during the Sep14 eruption

| Episode | Starting time (UT) | Starting frequency (MHz) | Harmonic present? | Ending time (UT) |
|---------|--------------------|--------------------------|-------------------|------------------|
| 1 | 11:00:34 | 45.1 | Yes | 11:05 |
| 2 | 11:12:53 | 32.9 | Yes | 11:16 |
| 3 | 11:14:24 | 25.6 | Yes | 11:16 |

### 3.3.1 NDA type II bursts from shock flanks

The onset of episode 1 at 11:00:34 roughly indicates the shock formation time. This is consistent with the EUV wave observed around 11:00:22 UT (Grechnev et al. 2018). From the height-time history of the shock in Fig. 6, we infer the shock height and speed to be 1.82 Rs and 1269 km/s at the time of episode 1. The burst had a clear fundamental and harmonic structure and well-defined starting frequency (45.1 MHz) of the fundamental. The sharp cutoff of the radio emission at 180 MHz suggests that plasma levels above 90 MHz are occulted. Therefore, the type II emission at 45 MHz is unocculted. According to Newkirk's density model above active regions, the density is given by n (cm$^{-3}$) = $n_0 10^{4.32/r}$, where r is the heliocentric distance in units of Rs, and $n_0 = 2\times10^4$. This distribution is equivalent to a power-law distribution n ~ $r^{-6.59}$, indicating a steep drop in density as a function of distance in the corona. Gopalswamy et al. (2013b) found a similar steep decline, $r^{-7.56}$, based on EUV observations in the early part of solar cycle 24. According to these models, the plasma level corresponding to the type II starting frequency of 45.1 MHz should be at a heliocentric distance of 1.75 and 1.66 Rs, respectively. These values are close to the measured shock height of 1.82 Rs using the model fit, suggesting that the type II source is likely to be within 15º - 24º from the shock nose. The drift rate (df/dt) of the type II burst was 0.13 MHz s$^{-1}$, which is typical of metric type II bursts (Mann et al. 1996; Gopalswamy et al. 2009b). The shock speed V can be derived from the drift rate using the relation,

V = 2L(1/f)(df/dt),    (1)

where L = |n(dn/dr)$^{-1}$| is the density (n) scale height and f is the emission frequency (equal to the local plasma frequency). For a power-law distribution of the form

n = $n_0 r^{-\alpha}$, with L = r/α,    (2)

so we get V= 1076 km/s. Taking into account of the possibility that the source may be slightly away from the nose, the speed becomes 1114 km/s, close to the local shock speed (1269 km/s). If the emission comes from the nose, the drift rate relation can be used with V = 1269 km/s to get L = 0.33 Rs. Since r = 1.82 Rs, we get α = 5.59, which is also reasonable ($n_0$ = 7.14×10$^8$ cm$^{-3}$).

The type II episodes F2 and F3 occur when the shock nose is in the height range 3.5 to 4 Rs, moving with a very high speed (>2000 km/s). The plasma frequencies at these heights are in the range 7.3 to 5 MHz. While DH type II emission is present at these frequencies, the type II emission at 32.9 and 25.6 MHz cannot come from the shock nose. According to the density distribution noted above (n ~ r$^{-5.59}$), 32.9 and 25.6 MHz plasma levels occur at distances 2.04 and 2.23 Rs, respectively. This means, these type II bursts come from distant flanks of the shock, about 56º away from the shock nose. The flank speed is expected to be >1000 km/s, which is high enough to support a shock. Note that different type II episodes might originate from different parts of the three-dimensional shock surface that cuts the appropriate plasma level. Episode 2 had an initial positive slope, which is likely due to the flank crossing a high-density structure such as a streamer. The Radio Solar Telescope Network (RSTN) reported a type II burst in the range 53-25 MHz during the interval 11:13 to 11:24 UT, which overlaps with episodes 2 and 3. Episode 1 was not reported by RSTN. The complex nature of the metric type II episode is typical of energetic eruptions in which many sections of the shock front is likely to produce type II bursts (see e.g., the 2017 September 10 eruption reported in Gopalswamy et al. 2018c).

**3.3.2 Heliocentric distance of the shock at SGRE end**

The IP type II burst can be used to determine the heliocentric distance of the shock until which it was accelerating >300 MeV protons in sufficient numbers to produce SGRE. For this we compute the drift rate of the type II burst over a stretch of the burst before the first break in the dynamic spectrum around 16:15 UT (see Fig. 7). Between 13:54 and 16:12 UT, the fundamental component of the burst drifts from 0.43 MHz to 0.31 MHz, yielding a drift rate of $1.35\times10^{-5}$ MHz s$^{-1}$. From the height-time plot of the shock in Fig. 6, the shock speed at 12:10 UT was ~2300 km/s and was slowly decreasing at the rate of 0.027 km s$^{-2}$. This average deceleration would bring down the shock speed to ~1900 km/s at 16:12 UT.

We can obtain the heliocentric distance over which the shock sends particles to the Sun to produce γ-rays. We select 16:12 UT as the reference time, which is close to the break in the type II burst noted above. At this time, V = 1900 km/s, L =r/2 (α =2 in the IP medium), f = 0.31 MHz, and df/dt = $1.35\times10^{-5}$ MHz s$^{-1}$. Substituting these quantities in eq. (1), we get r = 62.7 Rs at 16:12 UT. Another way is to uses the shock height measured from the GCS fit and extrapolate it to 16:12 UT using the density fall off (equation 2). From the height-time plot in Fig. 6 we see that at 12:10 UT, the shock was at a height of 15 Rs and the plasma frequency is ~1.2 MHz, which corresponds to a plasma density of $1.78 \times 10^4$ cm$^{-3}$. The local plasma density at 16:12 is given by the plasma frequency 0.31 MHz as 1186 cm$^{-3}$. If the density varies as r$^{-2}$, we can get the shock distance from the plasma densities at 12:10 UT and 16:12 UT as 58.1 Rs, which is similar to the one derived from the drift rate and differs only by ~7.3%. This result is consistent with the shock heliocentric distance of ~90 Rs at the end of the 2015 June 21 SGRE reported in Gopalswamy et al. (2018b).

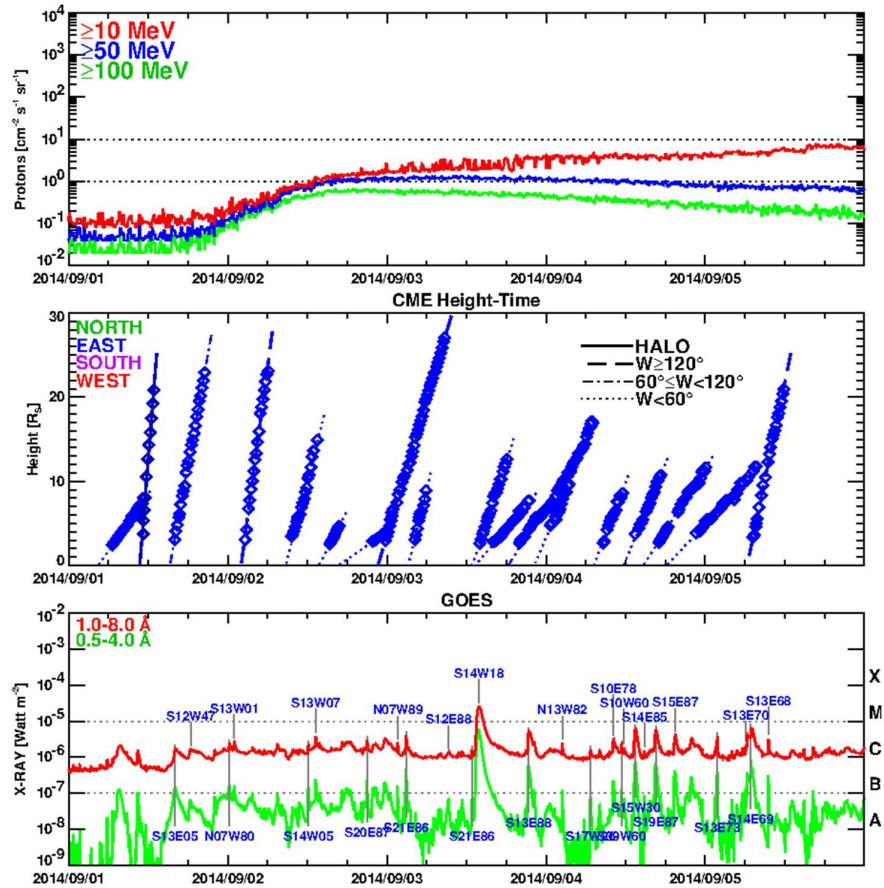

Figure 9. SEP, CME, and flare activities observed over the interval 2014 September 1-5 from the Sun-Earth line. (top) GOES protons flux in three integral energy channels, >10 MeV (red), >50 MeV (blue), and >100 MeV (green). (middle) CME height-time plots for CMEs heading in the east direction. (bottom) GOES soft X-ray intensity as a function of time. In the middle panel, only eastern CMEs are included. There was no CME from the western hemisphere with SEP association. Only the Sep14 CME was a halo CME.

**3.4 SEP association**

Figure 9 shows the proton intensity observed by GOES as a function of time over several days. The peak SEP flux in the >10 MeV energy channel is <10 pfu, making it a minor event at Earth. The onset of the SEP event is delayed by ~9 hrs because of the poor connectivity to Earth (the well-connected field lines are about 170º away from the source region). The SEP intensity is roughly the same in all three GOES energy channels (>10 MeV, >50 MeV, and >100 MeV) until about 16 UT on September 2, suggesting a hard spectrum. The intensity decayed extremely slowly in all three channels: until September 7 in the >100 MeV channel and until September 10 in the other two channels. A similar long-enduring GOES particle intensity was observed during the backside extreme event of 2012 July 23 (Gopalswamy et al. 2016). The height-time plots of all the CMEs are also shown in Fig. 9 (middle). It is clear that the only full halo CME is the CME in hand. None of the CMEs had any effect on the intensity curves, except the one on September 5. The GOES plot at the bottom indicates that the flare is occulted in the Sep14 event.

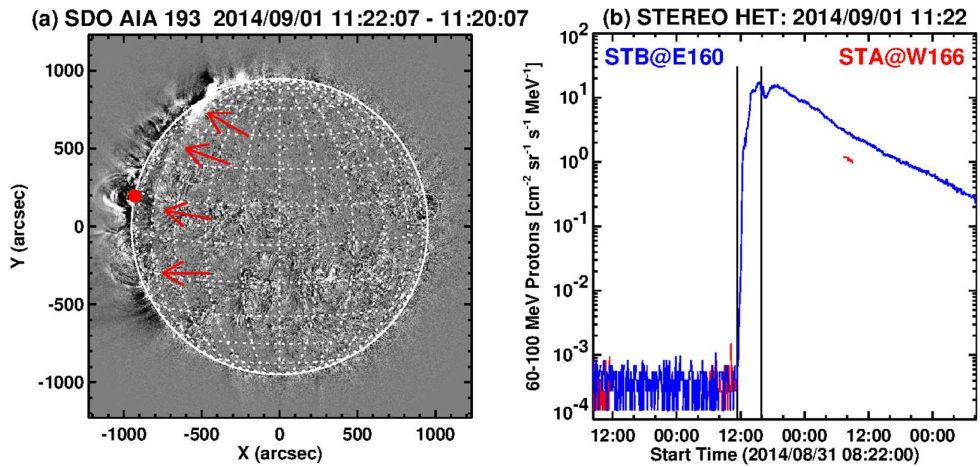

Figure 10. (a) EUV shock on the solar disk (along E60 marked by arrows) propagating away from the backside source as seen in a EUV difference image at 11:22:07 UT. The red dot at the limb is the location of SGRE at 11:15 UT revised from Ackermann et al. (2017). (b) Plot of proton flux in the 62-100 MeV energy channel of STB/HET (blue). STA had data gap except a brief moment (red). The first vertical line marks the time of the SDO/AIA image on left (11:22 UT). The second vertical line marks the end of SGRE (15:52 UT). The STA and STB were at W166 and E160, respectively from Earth view.

Figure 10a shows the revised SGRE source location along with the EUV wave front on the solar disk (pointed by arrows). Note that the EUV front corresponds to the ground track of the EUV wave dome represented by the spheroid in Fig. 5. Thus, the SGRE source location corresponds to the field lines located between the active region site and the shock flank. Since STB was well-connected to the source, an intense SEP event was detected by that spacecraft. The proton intensity in Fig. 10b in the 62-100 MeV energy channel sharply rises with the eruption and slowly decays over a few days, consistent with the GOES intensity profile. The SGRE event lasted roughly until the declining phase of the first peak in the 62-100 MeV intensity profile observed at STB. The particle event was also detected at STA, which was located at W166, but only briefly due to large data gaps. The intensity was consistent with the STB curve, although a bit lower during the brief period (centered at 8 UT on September 2).

### 3.4.1 Fluence spectrum

We now confirm that the Sep14 SEP event had a hard spectrum, using STEREO data (10-100 MeV), consistent with the spectrum obtained from the Payload for Matter-Antimatter Exploration and Light Nuclei Astrophysics (PAMELA) data at energies >80 MeV (Bruno et al. 2018) PAMELA detector recorded SEPs up to 700 MeV. De Nolfo et al. (2019) determined that the Sep14 event had event-integrated intensity at >500 MeV as $1.5\times10^3$ cm$^{-2}$ sr$^{-1}$. Figure 11a shows the SEP intensity at various STEREO energy channels and the fluence values in each of these channels. The event in question was preceded and followed by other SEP events, so the fluence computed is approximate, especially because it was difficult to decide the background level. However, choosing background levels at different intervals roughly resulted in the same fluence spectrum. The derived fluence spectrum is extremely hard, at the theoretic limit of 2. Figure 11b shows the spectrum with a fit to the 10-100 MeV data points from STEREO LET and HET. The spectral index is -1.81, close to the theoretical value for diffusive shock acceleration (see, e.g., Vainio, 2009; Wolff and Tautz 2015). The GOES HEPAD data also yields a hard spectrum with a spectral index of ~ -2.01, in good agreement with the STEREO data. The spectrum is thus one of the hardest, similar to well-connected GLE events. Cohen and Mewaldt (2018) found this event to be an extreme event with one of the hardest fluence spectrum. In our previous work (Gopalswamy et al. 2018a), we noted that the lack of >300 MeV particles at Earth in the 2011 March 07 SGRE event is due to poor latitudinal connectivity (the shock nose is too far above the ecliptic) because of the source location (N31) and an unfavorable solar B0 angle (-7º.25) (see Gopalswamy et al. 2013a; Gopalswamy and Mäkelä, 2014). The Sep14 event had the opposite situation: The source latitude was N16, while the B0 angle is +7º.25, so the effective source location with respect to the ecliptic was N09, well within the average nose distance of GLE events (~±13º). Accordingly, STB observed a large SEP event and the computed >100 MeV proton intensity remained high during the SGRE event. Even though the highest HET energy channel is 62-100 MeV, we infer that there were particles of significantly higher energy, possibly up to GeV energies.

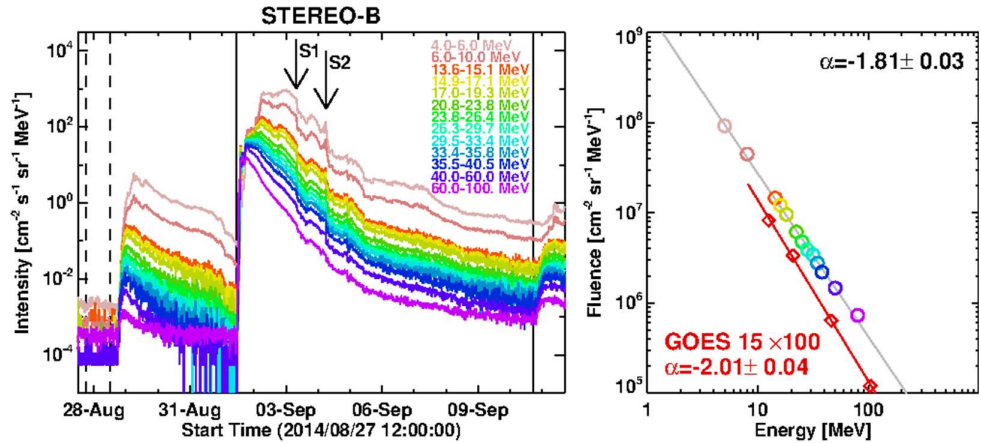

Figure 11. (left) Proton intensities at various energy channels from STB. The 4-6 MeV and 6-10 MeV data are from STB/LET and the remaining are from STB/HET. The fluences were determined over the interval between the vertical dark lines. The background intensity of protons was obtained as an average over the interval marked by the vertical dashed lines (from 18:00 UT on 2014 August 27 to 12:00 UT on 2014 August 28) to avoid preceding SEP event from a different source region. S1 (September 3 at 07:45 UT) and S2 (September 4 at 06:07 UT) are the shocks detected in situ at STB; S1 is associated with the CME in question, indicating a transit time of ~45 hr. (right) The fluence spectrum from STB. The spectrum is fit to a power law (gray line) with an index of 1.81±0.03. The red line is the spectrum obtained from GOES data. The GOES fluences are multiplied by a factor 100 and fit to a power law with an index of 2.01±0.04.

### 3.4.2 Event size at >10 MeV energies

The high intensity of the event at STB is evident from the "snow storm" visible in Figs. 3 and 4. In fact STB/COR1 images had the snow storm at least until the end of September 3 (see https://cdaw.gsfc.nasa.gov/stereo/daily_movies/2014/09/03/). We computed the >10 MeV intensity at STB from the spectrum in the 10-100 MeV range every 60 min under the assumption that there are no particles with energies >1 GeV. We average the data over 60 min because of the velocity dispersion, the spectrum was highly variable over shorter intervals. We use the following procedure to get the >10 MeV flux. STEREO/HET records data in the energy range 13.6 to 100 MeV in 11 energy channels. The observations need to be extended to 1000 MeV on the higher energy side and to 10 MeV on the lower energy side. First, we estimated the flux in the HET energy range 13.6 to 100 MeV by fitting a power law to the observations. Second, we obtain the 100-1000 MeV flux assuming the power-law index to be 1.81 (same as the power-law index of the fluence spectrum in Fig. 11b) that matches the flux in the 60-100 MeV channel until about 04:30 UT on September 2; beyond that, the actual spectral indices obtained from all HET channels are used. Third, we obtain the flux in the 10 to 13.6 MeV in two ways: (i) assuming the power law fitted to the HET data is valid down to 10 MeV, and (ii) interpolating between the highest energy channel in LET (6-10 MeV) and the lowest energy channel in HET (13.6-15.1 MeV). The background flux is subtracted in all cases. By adding the fluxes in the three steps above gives the >10 MeV flux. Figure 12 (left) shows the >10 MeV flux obtained. The flux obtained from HET data alone gives a >10 MeV peak flux of 4170 pfu. Combining the LET and HET data gives a >10 MeV peak flux of ~3564 pfu. The peak fluxes differ only by ~17%, suggesting that the Sep14 event is truly an intense event. The derived >10 MeV flux at STB is thus larger than that at GOES by more than two orders of magnitude.

Figure 12 (right) shows the time evolution of the >100 MeV STB proton flux computed by extrapolating the 10-100 MeV spectrum to higher energies assuming that the power-law index of the fluence spectrum is also applicable to 1-hour intervals and assuming that the number of particles with energy >1 GeV is negligible. The >100 MeV proton flux serves as a proxy to the >300 MeV proton flux required for SGRE. Also shown is the >100 MeV γ-ray flux from Fermi/LAT. The >100 MeV proton flux increased by four orders magnitude around the onset of SGRE and remained high throughout the event. The SGRE duration is longer than all impulsive phase activities and the end of soft X-ray flare emission and closer to the type II burst duration. The intensity of the flare in EUVI (STB), shown as a proxy to the soft X-ray flare emission coincides with the SGRE peak (within ~5 min time resolution of the EUVI observations). Share et al. (2018) used the soft X-ray count rates from the MESSENGER SAX instrument to identify the equivalent GOES flare duration between 10:56 UT and 11:34 UT, with a peak at 11:10 UT coincident with the STB/EUVI light curve.

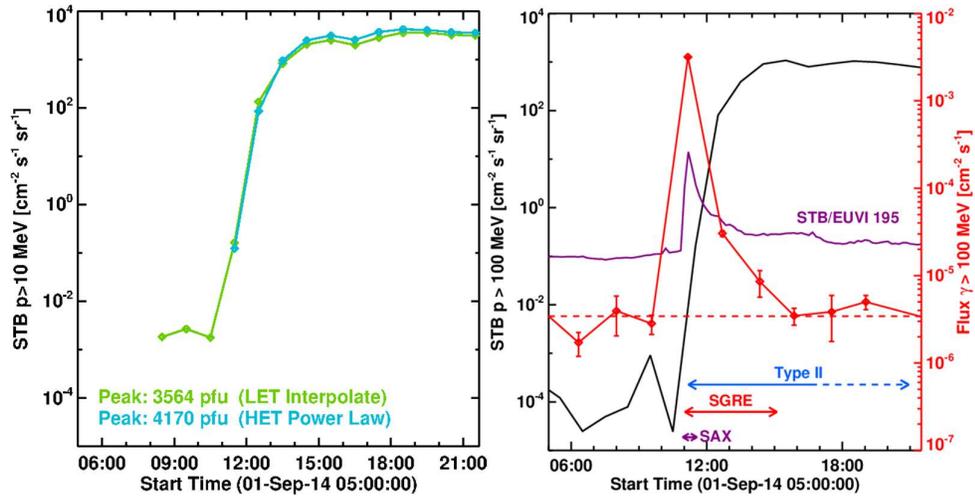

Figure 12. (left) Time profiles of the >10 MeV proton flux obtained from STB data with two different ways of estimating the flux between 10 and 13.6 MeV: (i) Using HET data alone by fitting a power law to the HET data and assuming that the power law is valid down to 10 MeV (red) and (ii) interpolating between the last (6-10 MeV) LET data point and first (13.6-15.1 MeV) HET data point (blue). The peak fluxes obtained by the two methods are shown on the plot. (right) Time profiles of the >100 MeV proton flux from STB (black curve), >100 MeV SGRE flux (red curve) and the flare represented by the STB/EUVI intensity (purple curve in arbitrary units). In obtaining the >100 MeV proton flux, we started from the 60-min averaged flux in the 60-100 MeV HET channel and used the power-law index of the fluence spectrum (1.81) to extend up to 1 GeV. The horizontal dashed red line marks the γ-ray background. The duration of the type II burst from SWAVES dynamic spectrum and the SGRE duration are marked by the blue and red double arrows, respectively. The purple double arrow marks the duration (10:55 UT to 11:34 UT) of the soft X-ray flare derived from the MESSENGER SAX instrument (from Share et al. 2018).

### 3.5 Duration comparisons

Gopalswamy et al. (2018a) reported a linear relationship between the durations of SGRE events ($T_{SGRE}$) and type II bursts ($T_{II}$) for a set of 13 SGRE events. They had also shown the 1991 June 11 SGRE event from Kanbach et al. (1993), but not included in the correlation. In Fig. 13, we show the $T_{SGRE}$ - $T_{II}$ plot that includes the 1991 June 11 event. The best fit to the scatter plot gives the following relation (all durations are in units of hours):

$T_{SGRE} = (1.0 \pm 0.2) T_{II} + (0.1 \pm 2.1)$.     (3)

This relation is not too different from the one in Gopalswamy et al. (2018a), confirming the linear relationship.

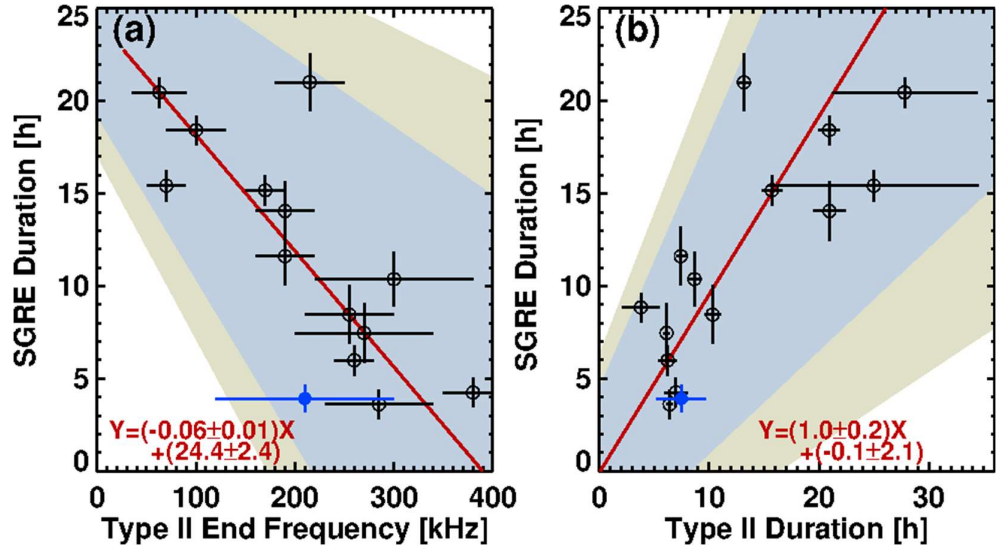

Figure 13. The SGRE duration, type-II ending frequency, and type II duration of the Sep14 event (blue data points) compared with 14 SGRE events reported in Gopalswamy et al. (2018a). The red lines in (a) and (b) are the best-fit lines for the 14 events. The blue and yellow shaded areas denote the 95% and 99% confidence intervals of the fit. The blue data points were not included in the fit but are consistent with the linear relationships in (a) and (b). Note that the Sep14 SGRE duration is expected to be a lower limit, so the agreement is likely better when the actual duration of SGRE is considered (the blue data point would move up toward the regression lines in (a) and (b).

We have over plotted the Sep14 data point in Fig. 13 with $T_{SGRE}$ = 3.92 ± 0.76 hr and $T_{II}$ = 7.5 ± 2.25 hr. The ending frequency of the type II burst is 210 ± 90 kHz. The Sep14 event agrees with both the best fit lines in Fig. 13. Note that we have used a generous error bar for the type II burst duration, taking the break at 16:18 UT and the end at 20:56 UT as possible end times. It must be noted that the SGRE duration given in Fig. 1 is likely to be an underestimate because we are not observing the entire flux of the γ-rays. There must be protons precipitating into other parts of the chromosphere that have magnetic connectivity to the shock nose. This limits the flux of gamma rays, and hence the last SGRE data point. Ackermann e al. (2017) compared the Sep14 event with disk events of similar flare size. They found that the total γ-ray energy is greater for on-disk flares probably due to the fact they are observed over longer timescales. Any increase in SGRE duration would move the data points in both plots in Fig.13 closer to the best fit lines. Thus, we can conclude that the Sep14 SGRE event is consistent with other SGRE events with a similar duration. Furthermore, Gopalswamy et al. (2019) extended the scatter plot to include SGRE events with duration exceeding 3 h. The resulting best-fit line is given by $T_{SGRE} = (0.9\pm0.2) T_{II} + (-0.8\pm1.9)$, which is not too different from the one given by eq. (3).

## 4. Discussion

The detailed investigation presented in this paper focused on how the SGRE event is related to the eruption geometry, 3-D kinematics of the CME and shock, SEP event, and various types of radio bursts. The type II burst extending to very low frequencies is a signature of a strong shock, which is confirmed by the ultra-fast CME (speed >2000 km/s). The SEP spectrum is very hard – typical of GLE events. The hard spectrum is confirmed using GOES and STB data and is consistent with the observation of high-energy particles by PAMELA (Bruno et al. 2018). Gopalswamy et al. (2016b) reported the fluence spectral indices of 86 western SEP events from solar cycles 23 and 24. The spectral index ranged from 2.01 to 6.12. The spectral index of only four events is similar to that of the Sep14 event (index <2.1): 1997 November 06 (2.07), 1998 May 02 (2.01), 2006 December 13 (2.07), and 2012 July 08 (2.01). The first three are GLE events and the last one is a large SEP event. A reanalysis of the last event showed that the spectral index is slightly larger, ~2.6. Thus, we are confident that the 2014 September 01 event is similar to GLE events indicating that >300 MeV particles required by the SGRE event are definitely present.

**4.1 The Magnetic Structure Associated with SGRE**

In this section, we develop a schematic model based on these observations and their synthesis into a coherent picture. In particular, we focus on how the temporally and spatially extended nature of the SGRE event follow naturally from the observations, especially when combined with the possible magnetic structures involved and how they are accessed by energetic particles accelerated at the flare site and the shock front. We attribute the temporally extended nature to the continued acceleration of high-energy particles by the CME-driven shock. The large spatial extent is determined by the footprint of the shock sheath that contains the open field lines surrounding the flux rope and threading through the shock nose. The SGRE source extension is naturally much larger than the post-eruption arcade to which the impulsive phase gamma-rays are confined.

One of the main characteristics of the SGRE events is that the γ-ray emission is temporally distinct from the impulsive-phase emission (Share et al. 2018). The 2014 September 01 event showed that the SGRE emission is likely to be spatially distinct from the impulsive phase emission. While the impulsive phase emission is confined to the post-eruption arcade, the extended phase γ-rays are associated with a structure much larger than the PEA.

A typical solar eruption involves two closed magnetic structures: the PEA and the flux rope, both of which are thought to be formed during the reconnection process. The PEA remains anchored to the solar surface, while the flux rope is ejected. The flux rope is also anchored to the solar surface, but the separation between the flux rope feet is larger than the extent of the PEA (see Fig. 14 and Webb et al. 2000; Gopalswamy 2009). The large extent of the flux rope is readily inferred from the fact that all SGRE events with duration >3 hr are associated with halo CMEs (Gopalswamy et al. 2018a; 2019) and halo CMEs are inferred to be fast and wide from quadrature observations (Gopalswamy et al. 2013c).

Particles accelerated during the reconnection process have access to both these structures. Electrons entering the PEA produce various microwave bursts and stationary type IV bursts (also known as flare continuum) in radio and hard X-ray bursts, while protons produce the impulsive phase γ-ray bursts. The spatial extent of the impulsive-phase emissions is defined by the size of the PEA. Electrons trapped in the flux rope produce moving type IV bursts. As the flux rope expands and moves away from the Sun, the electrons lose energy and the moving type IV burst decays. When the flux rope is super-Alfvenic, it drives a fast-mode MHD shock, which stands off from the flux rope at a distance determined by the radius of curvature of the flux rope and the shock Mach number (see, e.g., Gopalswamy and Yashiro 2011). Open magnetic field lines surrounding the flux rope thread through the shock and form the third magnetic structure of interest. The schematic in Fig. 14 shows these magnetic structures. The PEA with dimming regions has been observed during the 2015 June 21 SGRE event and a flux rope was reconstructed using white-light observations (Gopalswamy et al. 2018b, their figures 5 and 13).

The shock is another source of energetic particles that have access to the open filed lines threading through the shock and hence travel both toward and away from the Sun. The shock-accelerated particles thus occupy a volume much larger than that of the PEA and flux rope and surround the flux rope. Electrons escaping the shock front produce type II radio bursts at the local plasma frequency and harmonics. The bursts can start typically at a frequency of ~150 MHz and extend to lower frequencies, ending at different frequencies depending on the strength of the shock. Type II bursts extending to very low frequencies (~20 kHz) close to the local plasma frequency at the observing spacecraft indicate the strongest of shocks that also result in large SEP events (Gopalswamy 2006). Type II bursts are produced typically by ~10 keV electrons via the plasma emission mechanism, while higher-energy electrons are observed in-situ as electron events. Higher-energy electrons can also propagate toward the Sun and may contribute to bremsstrahlung continuum. Protons accelerated by the shock are detected in situ as SEP events, while those propagating toward the Sun produce γ-rays (SGRE) if the proton energy exceeds 300 MeV.

Figure 14 sketches the scenario of >300 MeV protons streaming toward the Sun and precipitation to produce SGRE. The protons are accelerated in the vicinity of the IP shock driven by the CME flux rope. It is implied that the legs of the CME flux rope are rooted on the Sun at the core-dimming regions on either side of the neutral line, but outside the post-eruption arcade. Open field lines threading the shock-nose intersect the solar surface just outside the flux rope. SGRE events require >300 MeV protons, which are generally accelerated closer to the nose, where the shock is the strongest. In this sense, the scenario depicted in Fig. 14 can be thought of as a variant of the cartoon model in Cliver et al. (1993), where protons from the shock flanks precipitate to the photosphere to produce the γ-ray line emission, which requires lower proton energies. The extent of the shock nose and its geometry is expected contribute significantly to the γ-ray variability in eruptive events. In particular, the orientation of the flux rope is likely to play a significant role in the varying shock-nose extent. Future γ-ray missions with a better spatial resolution should be able to discern the spatial distribution of γ-rays around the eruption region.

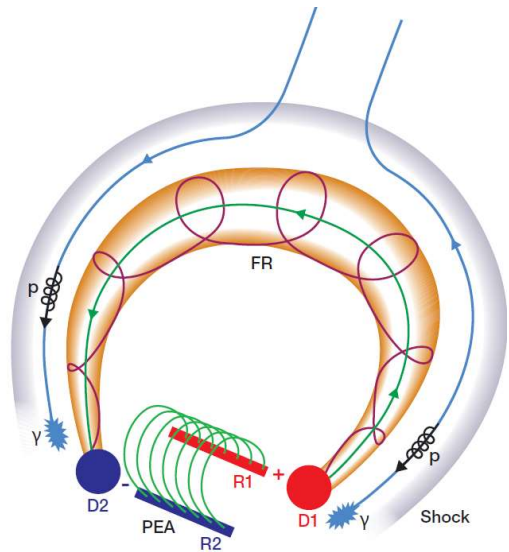

Figure 14. A schematic showing the magnetic structures involved in a typical eruption. The post-eruption arcade (PEA) with its feet rooted on the flare ribbons R1 and R2. The flux rope (FR) is a much larger structure rooted in the dimming regions D1 and D2. The flux rope is surrounded by a shock dome. Note that all these structures are directly inferred from observations during the 2014 September 01 event. Open magnetic field lines (represented by the blue lines) in the sheath region that thread through the shock nose are rooted in the region beyond the dimming regions. Protons (>300 MeV, black coils) accelerated at the shock front travel along these open field lines precipitate to the solar chromosphere (indicated by the patches at the bottom of the blue lines) and produce γ-rays; the separation between the patches represents the largest spatial extent of the SGRE source.

Recently Hudson (2018) criticized the CME-shock solution by suggesting that a strong mirror force prevents protons from propagating sufficiently close to the Sun. The mirroring is a common problem to particles precipitating along any magnetic structure: PEA, flux rope, or the open structures threading the shock nose. It must be noted that in the case of PEA, particles precipitate from high magnetic field regions to higher magnetic field regions. On the other hand, in the case of particles precipitating along open field lines, the precipitation occurs from low-field region ahead of the shock nose to the quiet-Sun field regions beyond the flux rope legs. Hudson (2018) suggested a "lasso" scenario, where accelerated protons are captured by closed magnetic fields that form a noose extending to heights of several solar radii. As the loop retracts in the aftermath of the CME, high-energy protons are transported by advection into denser solar atmosphere. In a typical eruption scenario, the flux rope structure is rapidly moving and expanding, while the PEA evolves slowly. It is not clear how the magnetic lasso is created and where it is placed with respect to the PEA, flux rope, and shock.

While the shock-sheath region naturally extends to the front-side allowing protons to precipitate and produce pion-decay γ-rays, Grechnev et al. (2018) invoke a static magnetic field structure emanating from the eruption site and ending on the frontside of the Sun. They show this using potential field extrapolation using a pre-event synoptic magnetic chart. Photospheric magnetic fields seldom show changes due to eruption. Non-potential fields possess the energy for the eruption, so potential field extrapolation cannot explain the eruption-associated disturbances in a violent event such as the Sep14 event. EUV images show another active region between the solar limb and the region of interest, so it is not clear if the structure emanated from the flaring region. Such a structure also does not fit in the clear PEA-flux rope structure observed and depicted in Figs. 3-5. Therefore, it is not clear if such large structures may explain particle acceleration and trapping within such large coronal structures not causally connected to the CME shock trap and accelerate particles to produce the observed SGRE (de Nolfo et al. 2019a,b).

The hard X-ray data presented by Grechnev et al. (2018) has a weak emission in the energy range 161-195 KeV, which may be from the same population of electrons that produced the gyrosynchrotron emission peaking at ~1000 MHz. Using a parametric fit to the radio flux density spectrum and analysis of X-ray emissions detected by Fermi's Gamma-ray Burst Monitor (GBM) Carley et al. (2017) inferred a common origin of non-thermal electrons responsible for the radio and X-ray emissions. The duration of the weak coronal hard X-ray emission is similar to that of the gyrosynchrotron sources and that of the DH type III burst. These emissions suggest that the flare-accelerated electrons are injected into the flux rope producing the hard X-ray and radio emissions. Particles from the same source but injected into the post-eruption arcade produce impulsive phase emissions and the flare continuum at radio wavelengths. Particles accelerated at the shock do not have access to these two magnetic structures (PEA and flux rope), but to the field lines surrounding the flux rope.

## 4.2 Proton numbers derived from SEP events and SGRE

We noted earlier that the gamma-ray flux in the Sep14 event is expected to be a lower limit because only that part of the emission originating from the frontside of the Sun is recorded by Fermi/LAT. Gamma-rays produced at other precipitation sites are occulted. Ackermann et al. (2017) suggested that the energy in the >500 MeV protons might be underestimated in events that occur at heliocentric angles >75º. Given the large size of the CMEs involved in SGRE events (half angle expected in the range 45º-50º), it is likely that gamma-ray flux in events originating within 30º from the limb are likely to be underestimated. Share et al. estimated the number of >500 MeV protons ($N_\gamma$) needed to produce SGRE as $1.99 \times 10^{30}$. De Nolfo et al. (2019a,b) reported the number of >500 MeV protons ($N_{SEP}$) observed in space by PAMELA to be two orders of magnitude higher: $2.35 \times 10^{32}$. Since $N_\gamma$ is likely to be underestimated for a limb/backside events, the difference between $N_\gamma$ and $N_{SEP}$ is expected to be much smaller for the Sep14 event. Using a sample of 14 events, De Nolfo et al. (2019a,b) concluded that $N_\gamma$ and $N_{SEP}$ are uncorrelated and that the γ-ray emission is probably not due to protons diffusing back to the Sun from CME-driven shocks. . However, if one accounts for the lack of nose connectivity in high-latitude events and the underestimate of $N_\gamma$, the correlation improves. A detailed report on the relation between $N_\gamma$ and $N_{SEP}$ taking into account of the connectivity issues in SEP events and the underestimate of $N_\gamma$ in limb events will be reported elsewhere.

Ackermann et al. (2017) reported that the total energy in >100 MeV γ-rays is about two orders of magnitude smaller than that in >500 MeV particles. For the Sep14 event, the energy in the >500 MeV protons is 50 times larger than that in the γ-rays ($1.4\times10^{24}$ vs. $7.0\times10^{25}$ erg). On the other hand, the energy in >500 MeV protons is seven orders of magnitude smaller than the CME kinetic energy. Using the observed CME mass of ~$1.6\times10^{16}$ g (https://cdaw.gsfc.nasa.gov/CME_list/UNIVERSAL/2014_09/univ2014_09.html) and the average speed in the coronagraph FOV of ~2000 km/s (see Fig.5), we get a kinetic energy of $3.2\times10^{32}$ erg. Mewaldt et al. (2005) have shown that CME-shocks are as efficient as supernova shocks in accelerating particles in that ~10% of CME kinetic energy goes into particle energy. It is clear that only a tiny fraction of the CME kinetic energy needs to go into the energy of >500 MeV protons to sustain the γ-rays. Furthermore, some numerical simulations and analytical work indicate significant excess of shock-injected protons as compared to the number of protons derived from SGRE observations (Kocharov et al. 2015; Afanasiev et al. 2018).

## 5. Summary and Conclusions

We performed a detailed investigation of the SOL2014-09-01 behind-the-limb eruption that resulted in the fourth largest SGRE event (in >100 MeV fluence) observed by Fermi/LAT. The fluence is likely a lower-limit to the actual fluence because some γ-ray photons could not have reached Fermi/LAT. The eruption produced a twin dimming and a post-eruption arcade and a flux rope rooted in the dimming regions. The low-inclination flux rope implies an eruption with a large east-west extension. The flux rope had a three-dimensional speed exceeding 2000 km/s and was driving a fast shock. The shock-accelerated particles had a hard spectrum similar to previous GLE events. The shock was also the source of nonthermal electrons that produced a type II burst with emission at frequencies ranging from tens of MHz to tens of kHz suggesting an intense shock propagating in the interplanetary medium. The SGRE, CME, and radio burst properties of this event are consistent with all the disk events that had SGRE durations ≥3 hrs. The main conclusions of the investigation are as follows.

1. The SGRE and type II burst durations in the SOL2014-09-01 backside eruption are consistent with the linear relationship between the durations found in all >3-hr SGRE events. The SGRE duration and the ending frequency of the type II bursts are also consistent with an inverse relationship between the two quantities obtained before.

2. The magnetic structures in the eruption (post eruption arcade, flux rope, shock, and shock sheath) inferred from observations suggest a large east-west extension making it possible for the presence of the particle-precipitation region on the frontside solar disk in Earth view.

3. The largest extent of the SGRE source is inferred as the separation between the shock sheaths at the flanks of the flux rope. This means if limb is within the face-on angular half width of a behind the limb flux rope, one can still see γ-rays on the frontside.

4. The large extent of the SGRE source implies that the part of it is hidden in the case of limb events.

5. The CME/shock kinematics, the hard SEP spectrum, and the association with type II radio bursts over a wide range of wavelengths imply a very energetic CME with copious production of high-energy (up to GeV) particles needed for the production of SGRE via neutral pion decay. This is consistent with the fact that some >1 GeV γ-ray photons were observed that require multi-GeV protons.

6. The SGRE ended when the shock reached a heliocentric distance of ~60 Rs, consistent with previous studies.

7. The equivalent flux of >10 MeV SEPs computed from STEREO particle spectra, is >3000 pfu, which makes it one of the largest SEP events in solar cycle 24.

**Acknowledgements**

This work benefited from NASA's open data policy in using Fermi, Wind, SOHO, SDO, and STEREO data. We thank NOAA for making GOES X-ray and particle data available online. SOHO is a project of international collaboration between ESA and NASA. STEREO is a mission in NASA's Solar Terrestrial Probes program. CALLISTO data are made available online at e-callisto.org by the Institute for Data Science FHNW Brugg/Windisch, Switzerland (PI: C. Monstein). Orfées is part of the FEDOME project, partly funded by the French Ministry of Defense. The We thank the Nançay Radio Observatory for making NDA data available online at http://www.obsnancay.fr. Work supported by NASA's Living with a Star program. HX was partially supported by NASA HGI grant NNX17AC47G. We thank the anonymous referee for helpful comments.

Forrest, D.J., Vestrand, W.T., Chupp, E.L., Rieger, E., Cooper, J.F., Share, G.H.: 1985, Neutral Pion Production in Solar Flares, *19th International Cosmic Ray Conf.* **4**, 146. ADS.

Gonzalez, W. D., Clúa De Gonzalez, A. L., Dal Lago, A., Tsurutani, B. T., Arballo, J. K., Lakhina, G. S., Buti, B., and Ho, G. M.: 1998, Magnetic cloud field intensities and solar wind velocities, *Geophys. Res. Lett.*, **25**, 963. DOI. ADS

Gopalswamy, N.: 2006, Coronal Mass Ejections and Type II Radio Bursts. In: N. Gopalswamy, R. Mewaldt, and J. Torsti (ed.), Solar Eruptions and Energetic Particles. *Geophysical Monograph Ser.* **165**, American Geophysical Union Press, 207. DOI. ADS.

Gopalswamy, N.: 2009, Coronal mass ejections and space weather. In: T., Tsuda, R., Fujii, K., Shibata, and M.A., Geller (ed.), *Climate and Weather of the Sun-Earth System (CAWSES): Selected Papers 2007 Kyoto Sympo.*, Terrapub, Tokyo, 77. DOI. ADS.

Gopalswamy, N.: 2011, Coronal mass ejections and solar radio emissions. In: H.O. Rucker, W.S. Kurth, P. Louarn, G. Fischer (ed.), *Planetary, Solar and Heliospheric Radio Emissions VII*, Austrian Academy of Sci. Press, 325. DOI. ADS.

Gopalswamy, N., Mäkelä, P.: 2014, Latitudinal Connectivity of Ground Level Enhancement Events, In: Q. Hu and G.P. Zank (ed.), *Coronal Heating to the Edge of the Heliosphere. ASP Conf. Ser*. **484**, 63. ADS.

Gopalswamy, N., Yashiro, S.: 2011, The Strength and Radial Profile of the Coronal Magnetic Field from the Standoff Distance of a Coronal Mass Ejection-driven Shock. *Astrophys. J.* **736**, L11. DOI. ADS.

Gopalswamy, N., Yashiro, S., Akiyama, S., Mäkelä, P., Xie, H., Kaiser, M.L., Howard, R., Bougeret, J.-L.: 2008, Coronal Mass Ejections, Type II Radio Bursts, and Solar Energetic Particle Events in the SOHO Era. *Anna. Geophysicae*. **26**, 3033. DOI. ADS.

Gopalswamy, N., Yashiro, S., Michalek, G., Stenborg, G., Vourlidas, A., Freeland, S., Howard, R.: 2009a, The SOHO/LASCO CME Catalog. *Earth Moon and Planets*. **104**, 295. DOI. ADS.

Gopalswamy, N., Thompson, W.T., Davila, J.M., Kaiser, M.L., Yashiro, S., Mäkelä, P., Michalek, G., Bougeret, J.-L., Howard, R.A.: 2009b, Relation
34